\documentstyle[11pt,epsfig]{article}
\begin{document}
\title{{\bf Screening Length in ${\bf 2+1}$-dimensional Abelian
Chern-Simons  Theories}}
\author{V. S. Alves\footnote{Permanent address: Departamento de
f\'{\i}sica, Universidade Federal do Par\'{a}, 66075-110 Bel\'{e}m,
Brasil} , Ashok Das and Silvana Perez$^{*}$\\
Department of Physics and Astronomy,\\
University of Rochester,\\
Rochester, NY 14627-0171\\
USA}

\date{}
\maketitle

\begin{center}
{ \bf Abstract}
\end{center}

In this paper, we systematically study the question of screening
length in Abelian Chern-Simons theories. In the Abelian Higgs theory,
where there are two massive poles in the gauge propagator at the tree
level,  we show that the coefficient of one of them becomes negligible
at high temperature and that the screening length is dominantly
determined by the parity violating part of the self-energy. In this
theory, static magnetic fields are screened. In the fermion theory, on
the other hand, the parity conserving part of the self-energy
determines the screening length and  static magnetic fields are
not screened. Several other interesting features are also discussed.

\newpage

\section{Introduction}

The properties of a charged plasma, at finite temperature, have been
studied extensively in the past in $3+1$ dimensions
\cite{fradkin,kislinger,weldon,blaizot}  and it is known
that there are several interesting
features that emerge in thermal QED. For example, it is known that the
photon becomes massive at finite temperature, much like a particle
moving in a medium. Furthermore, since thermal amplitudes are, in
general, non-analytic at the origin in the energy-momentum plane
\cite{fetter-walecka,weldon1}, the
mass of the photon that manifests in different processes is
distinct. For example, the screening length between two static charges
is related to the electric mass of the photon, $m_{el}$, while the
length associated with plasma oscillations due to a sudden excitation
of the plasma is related to the plasmon mass, $m_{pl}$, and the two
masses are quite distinct. In $3+1$ dimensional QED, for example, the
electric mass is defined as 

\begin{equation}\label{eq1}
\lim_{\vec{p}\rightarrow 0} \Pi^{00} ( p^0=0, \vec{p}) = m_{el}^2\,,
\end{equation}
where $\Pi^{\mu \nu}$ denotes the photon self-energy and the potential
between two static charges separated by a distance $R=|\vec{R}|$ is
obtained to be 

\begin{equation}\label{eq2}
\sim\int \frac{d^3p}{(2 \pi)^3} D_{00} ( p^0=0, \vec{p}) e^{i
\vec{p} \cdot \vec{R}}= \int \frac{d^3p}{(2 \pi)^3} \frac{ e^{i
\vec{p} \cdot \vec{R}} }{\vec{p}^2 + m_{el}^2} = \frac{e^{-m_{el}
R}}{4 \pi R}\;. 
\end{equation}
This shows that the screening length and the electric mass are
inversely related. 

In $2+1$ dimensional QED, if we naively carry over the definition of
the  electric mass as in (\ref{eq1}) (as well as the propagator),
then,  although the potential between two static
charges would not have the form of a Yukawa potential as in
(\ref{eq2}),  it can be determined to be

\begin{equation}\label{eq3}
\sim  \int \frac{d^2p}{(2 \pi)^2} D_{00} ( p^0 = 0, \vec{p} )
e^{i \vec{p} \cdot \vec{R}}=\int \frac{d^2p}{(2 \pi)^2} \frac{
e^{i \vec{p}\cdot \vec{R}} }{\vec{p}^2 + m_{el}^2} = \frac{1}{2
\pi} K_0 (m_{el} R)\;. 
\end{equation}
Here $K_0(m_{el} R)$ is the Bessel function with the asymptotic behavior
(for large z) 

\begin{equation}
K_0(z) \longrightarrow \sqrt{\frac{\pi}{2 z}} e^{-z}\,,
\end{equation}
so that once again, we see that the screening length and the electric
mass are inversely related. 

In $2+1$ dimensions, however, we can also add a parity violating
Chern-Simons term to the gauge Lagrangian \cite{deser} and we know
that, in  some theories,  such a term 
can be generated through quantum corrections even if it is not present
at the tree level \cite{babu}. In such a case, we expect that $\Pi^{00}$ alone
cannot determine the electric mass which can, in principle, depend on the
Chern-Simons coefficient. Furthermore, in a $2+1$ dimensional Abelian
Higgs model with a Chern-Simons term \cite{khlebnikov,dunne,khare,alves}, it
is  known that even
the tree level gauge boson propagator has two distinct poles. This
raises the interesting question, namely, whether there is a unique
screening length in such theories. In this paper, we study this
question in detail in various Abelian Chern-Simons theories, which
leads to some interesting results. We note here that the question of
screening length, in a Yang-Mills-Chern-Simons theory interacting with
fermions, has been discussed in the past \cite{pisarski} and we
compare our  results with these.

\section{Abelian Higgs model with a Chern-Simons term}

Let us start with an Abelian gauge field, $A_{\mu}$, in $2+1$ dimensions
with both a Maxwell and a Chern-Simons term interacting with a charged
scalar field with a symmetry breaking quartic potential
\cite{khlebnikov,dunne,khare,alves}, 

\begin{equation}
{\cal L} = - \frac{1}{4} F_{\mu \nu} F^{\mu \nu} + \frac{\kappa}{2}
\epsilon^{\mu \nu \lambda} A_{\mu}\partial_{\nu} A_{\lambda} +
(D_{\mu} \Phi)^* (D^{\mu} \Phi) - \frac{\lambda}{4} ( \Phi^* \Phi -
v^2)^2 
\end{equation}
where $\kappa$ represents the Chern-Simons coefficient.

In the spontaneously broken phase, where $\Phi$ has a nonzero vacuum
expectation value, $\langle\Phi\rangle = v$, we expand the scalar
field as $\Phi = v + \frac{1}{\sqrt{2}} (\sigma + i \chi)$ to
obtain 

\begin{eqnarray}
{\cal L} &=&  - \frac{1}{4} F_{\mu \nu} F^{\mu \nu} + \frac{\kappa}{2}
\epsilon^{\mu \nu \lambda} A_{\mu}\partial_{\nu} A_{\lambda} +
\frac{m^2}{2} A_{\mu} A^{\mu} + {1\over 2} \partial_{\mu}\sigma
\partial^{\mu}\sigma + {1\over 2} \partial_{\mu}\chi
\partial^{\mu}\chi - m \partial^{\mu}\chi A_{\mu}
\nonumber \\
& & - e (\sigma\partial^{\mu}\chi - \chi\partial^{\mu}\sigma) A_{\mu} 
+ \frac{e^2}{2} ( \sigma^2 + \chi^2 + 2 \sqrt{2} v \sigma)
A_{\mu}A^{\mu} - \frac{\lambda}{16}(\sigma^2 + \chi^2 + 2 \sqrt{2} v
\sigma)^2, 
\end{eqnarray}
where we have defined
\begin{equation}
m= \sqrt{2} \,e v\;.
\end{equation}
We can add to this a gauge fixing Lagrangian as well as the
corresponding ghost Lagrangian. Let us note here that in the $R_{\xi}$
gauge the $A_{\mu}$-$\chi$ mixing term disappears and all the fields
have nontrivial mass parameters 
\begin{equation}
m^2 = 2 e^2 v^2,\qquad 
m_{\sigma}^2 = \lambda v^2,\qquad 
m_{\chi}^2 = m_{c}^{2} = \xi m^2,
\end{equation}
where $\xi$ is the gauge fixing parameter. 

At zero temperature, the tree level gauge propagator, in this case, has the
form 

\begin{equation}
D^{(0)}_{\mu \nu} (p) = - \frac{1}{(p^2 - m_+^2)(p^2 - m_-^2)} \left[
\eta_{\mu \nu} \,( p^2 - m^2) - p_{\mu}p_{\nu}
\frac{(1-\xi)(p^2-m^2) + \xi \kappa^2}{p^2 - \xi m^2} + i
\,\kappa \,\epsilon_{\mu \nu \lambda} \,p^{\lambda}\right], 
\end{equation}
where the superscript ``$0$'' denotes the tree level propagator and 
\begin{equation}
m_{\pm}^2 = \frac{\kappa^2 + 2 m^2 \pm (\kappa^4 + 4 \,m^2
\,\kappa^2)^{1/2} }{2}. 
\end{equation}
The two distinct poles of the propagator, alluded to in the
introduction,  are manifest in this case.

For the purpose of studying the gauge boson self-energy, it is much
more convenient for us to work in the unitary gauge, $\chi=0$, where
the number of relevant Feynman graphs is much smaller. At zero
temperature, the tree level gauge and the scalar propagators, in the unitary
gauge, have the forms 

\begin{eqnarray}
D^{(0)}_{\mu \nu} (p) & = & - \frac{1}{(p^2 - m_+^2)(p^2 - m_-^2)} \left[
\eta_{\mu \nu}\,(p^2 - m^2) - p_{\mu}p_{\nu} \frac{p^2 - m^2 -
\kappa^2}{m^2} + i\, \kappa \,\epsilon_{\mu \nu \lambda}
\,p^{\lambda} \right],\nonumber\\
D^{(0)}_{\sigma} (p) & = & \frac{1}{p^2 - m_{\sigma}^2},
\end{eqnarray}
where, again, the two poles in  the gauge boson propagator
are manifest.

We would like to study, systematically, the question of the screening
length in this theory at finite temperature before turning to the
fermion theory later. To make the problem
precise, let us note that we will work in the imaginary time formalism
\cite{kapusta,lebellac,das} 
where the tree level propagators take the forms 

\begin{eqnarray}
D^{(0)}_{\mu \nu}(p) & = &  \frac{1}{(p^2 + m_+^2)(p^2 +
m_-^2)} \left[ \delta_{\mu \nu}\,(p^2 + m^2) + p_{\mu}\,p_{\nu}
\frac{p^2 + m^2 + \kappa^2}{m^2} - \kappa \,\epsilon_{\mu \nu \lambda}
\,p_{\lambda} \right],\nonumber\\
D^{(0)}_{\sigma} (p) & = & \frac{1}{p^2 + m_{\sigma}^2},
\end{eqnarray}
with $p^0=\frac{2 n \pi}{\beta}$, $\beta= \frac{1}{T}$
and the Boltzmann constant $k=1$. Let $u_{\mu}$ denote the velocity
of the heat bath with $u_{\mu}u_{\mu}=1$. In the rest frame of the
heat bath, $u_{\mu}=(1,0,0,0)$. Let us also define \cite{weldon,das} (all of our
discussion is in the imaginary time formalism and, therefore, in
Euclidean space.) 
\begin{equation}
\bar{u}_{\mu} = u_{\mu} - \frac{u \cdot p}{p^2}\,p_{\mu},\quad
\tilde{p}_{\mu} = p_{\mu} - (u \cdot p)\, u_{\mu},\quad
\tilde{\delta}_{\mu\nu} = \delta_{\mu\nu} - u_{\mu}\,u_{\nu},
\end{equation}
which satisfy
\begin{equation}
p \cdot \bar{u} = 0 = u \cdot \tilde{p} = u_{\mu} \tilde{\delta}_{\mu \nu}\,.
\end{equation}

With these structures, let us define
\begin{equation}
P_{\mu\nu} = \tilde{\delta}_{\mu \nu} - \frac{\tilde{p}_{\mu}
\tilde{p}_{\nu}}{\tilde{p}^2},\qquad 
Q_{\mu \nu}=\frac{p^2}{\tilde{p}^2} \,\bar{u}_{\mu}\bar{u}_{\nu}.
\end{equation}
It can be easily checked that
\begin{eqnarray}
p_{\mu}\,P_{\mu \nu} & = & 0 \,=\, p_{\mu} Q_{\mu \nu} =
P_{\mu\nu}Q_{\nu\lambda},\nonumber\\
P_{\mu \nu} P_{\nu \lambda} & = & P_{\mu \lambda},\quad
Q_{\mu \nu} Q_{\nu \lambda} = Q_{\mu \lambda},\nonumber\\
P_{\mu \nu} + Q_{\mu \nu} & = & \delta_{\mu \nu} - \frac{p_{\mu}
p_{\nu}}{p^2}.
\end{eqnarray}

The self-energy for the gauge boson, at finite temperature, can now be
parameterized, to all orders, in the unitary gauge as 
\begin{eqnarray}
\Pi_{\mu \nu} (p) &=& P_{\mu \nu} \,\Pi_1 + Q_{\mu \nu} \,\Pi_2 +
\delta_{\mu \nu} \,\Pi_3 + \epsilon_{\mu \nu \lambda}\, p_{\lambda}\,
\Pi_{\rm odd}\nonumber \\ 
&=& P_{\mu \nu}\, (\Pi_1 + \Pi_3) +  Q_{\mu \nu}\, (\Pi_2 + \Pi_3) +
\frac{p_{\mu} p_{\nu}}{p^2} \,\Pi_3 +  \epsilon_{\mu \nu \lambda}\,
p_{\lambda}\, \Pi_{\rm odd}. \label{selfenergy}
\end{eqnarray}
Adding the tree level term, the complete two point function has the form
\begin{eqnarray}
\Gamma_{\mu \nu} (p) &=& P_{\mu \nu} \,(p^2 + m^2 + \Pi_1 + \Pi_3) +
Q_{\mu \nu} \,(p^2 + m^2 + \Pi_2 + \Pi_3)\nonumber\\
 &  & \quad
+ \frac{p_{\mu} p_{\nu}}{p^2}\,(m^2 + \Pi_3) + \epsilon_{\mu \nu
\lambda}\, p_{\lambda} \,( \kappa + \Pi_{\rm odd})\nonumber \\ 
&=& P_{\mu \nu} \,(p^2 + M_1^2) + Q_{\mu \nu} \,(p^2 + M_2^2) +
\frac{p_{\mu} p_{\nu}}{p^2}\,(m^2 + \Pi_3)
+ \epsilon_{\mu \nu \lambda} \,p_{\lambda} \,( \kappa + \Pi_{\rm odd}),
\end{eqnarray}
where we have defined
\begin{equation}
M_1^2 = m^2 + \Pi_1 + \Pi_3\,,\qquad
M_2^2 = m^2 + \Pi_2 + \Pi_3\,.
\end{equation}
The complete propagator, which is the inverse of the complete two
point function, can now be determined to be 
\begin{eqnarray}
D_{\mu \nu} (p) &=& \frac{1}{(p^2+M_+^2)(p^2 + M_-^2)}\left[P_{\mu
\nu} \,(p^2 + M_2^2) + Q_{\mu \nu} \,( p^2 + M_1^2) - (\kappa +
\Pi_{\rm odd}) \,\epsilon_{\mu \nu \lambda} \,p_{\lambda}\right] 
\nonumber \\
& &\quad +  \frac{p_{\mu} p_{\nu}}{p^2}\,\frac{1}{m^2 + \Pi_3},
\end{eqnarray}
where we have defined
\begin{equation}
M_{\pm}^2 = \frac{(\kappa + \Pi_{\rm odd})^2 + M_1^2 + M_2^2 \pm ((M_1^2
-M_2^2)^2 + 2 (M_1^2 + M_2^2)(\kappa + \Pi_{\rm odd})^2 + (\kappa +
\Pi_{\rm odd})^4)^{1/2}}{2}. \label{masses}
\end{equation}

There are several things to note from this structure. First of all,
the propagator continues to have two distinct poles. Second, the poles
of the propagator correspond to the mass scales $M_{\pm}$ which
involve the Chern-Simons term (with radiative corrections)
non-trivially so that it is not possible to identify the electric mass
with $\Pi_{00}$ as in (\ref{eq1}). Finally, let us note that we can rewrite
the propagator also as 
\begin{eqnarray}
D_{\mu \nu} &=& \frac{P_{\mu \nu}}{M_+^2 - M_-^2}\left[\frac{M_+^2 -
M_2^2}{p^2 + M_+^2} - \frac{M_-^2 - M_2^2}{p^2 + M_-^2} \right] 
+ \frac{Q_{\mu \nu}}{M_+^2 - M_-^2}\left[\frac{M_+^2 - M_1^2}{p^2 +
M_+^2} - \frac{M_-^2 - M_1^2}{p^2 + M_-^2} \right]\nonumber \\ 
& &\quad + \frac{\kappa + \Pi_{\rm odd}}{M_+^2 - M_-^2} \epsilon_{\mu \nu
\lambda}p_{\lambda} \left[\frac{1}{p^2 + M_+^2} - \frac{1}{p^2 +
M_-^2} \right] + \frac{p_{\mu} p_{\nu}}{p^2}\frac{1}{m^2 +
\Pi_3}. \label{poles}
\end{eqnarray}
Each tensor structure now has a sum of two simple poles and the problem
of the uniqueness of a screening length is now clear. In fact,
depending on the parameters of the theory, we note from eq. (\ref{poles}) that
a screening potential can even become an anti-screening potential. 
Of course, this has been a general analysis so far and only an actual
calculation can determine what really happens. 

Before presenting the actual calculations, let us note here that
although our discussion has so far been within the context of the 
Abelian Chern-Simons Higgs system, the same general features arise in the
$2+1$ dimensional QED with a Chern-Simons term, as we will discuss in
section {\bf 4}. Interestingly, although the answer to the uniqueness of the
screening length is similar in the two theories, the mechanisms
responsible for this are quite different, as we will see.

\section{The calculations}

\begin{figure}
\centerline{\epsfig{file=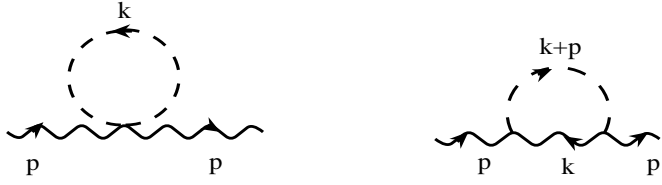, width = 14cm, height = 9cm}}
\vskip -4cm
\caption{Tadpole and Rising Sun Diagrams}
\end{figure}

In the unitary gauge, there are only two diagrams which contribute to
the photon self-energy (see figure 1). Of the two diagrams, it is only
the rising sun diagram which
contributes to the parity violating part of the photon self-energy which
has already been
calculated in \cite{alves}. Therefore, we will concentrate only on the parity
conserving part of these diagrams at finite temperature. Furthermore,
the tadpole diagram is independent of the external momentum and
consequently gives an analytic contribution -- it has the same value
both in the static as well as the long wave limits. The only
non-analyticity may possibly arise from the rising sun diagram. 

Before evaluating the individual diagrams, let us note that our
interest lies in calculating the form factors $\Pi_1$, $\Pi_2$ and
$\Pi_3$ ($\Pi_{\rm odd}$ has already been calculated in \cite{alves}). From the
parameterization of the self-energy in (\ref{selfenergy}), we note
that these  can
be determined from the self-energy as $(i,j=1,2)$ 
\begin{eqnarray}
\Pi_1 & = & - \frac{p_0^2}{p_0^2 - \vec{p}^2}\,\Pi_{00} + \delta_{ij}
\,\Pi_{ij} - \frac{p_0^2 - 2 \vec{p}^2}{p_0^2 - \vec{p}^2} \,\frac{p_i
p_j}{\vec{p}^2}\,\Pi_{ij},\nonumber\\ 
\Pi_2 & = & \frac{p_0^2 + \vec{p}^2}{p_0^2 - \vec{p}^2}\,( - \Pi_{00} +
\frac{p_i p_j}{\vec{p}^2}\, \Pi_{ij}),\nonumber\\ 
\Pi_3 & = & \frac{p_0^2}{p_0^2 - \vec{p}^2}\,\Pi_{00} -
\frac{\vec{p}^2}{p_0^2 -\vec{p}^2}\, \frac{p_i
p_j}{\vec{p}^2}\,\Pi_{ij}.\label{formfactors} 
\end{eqnarray}
Therefore, rather than calculating the self-energy, it is simpler to
calculate $\Pi_{00}$, $\delta_{ij} \Pi_{ij}$ and $\frac{p_i
p_j}{\vec{p}^2}\Pi_{ij}$ from which the three quantities of interest
can be determined. We note that, in the static limit $(p_0=0)$,
eq. (\ref{formfactors}) leads to 
\begin{eqnarray}
\Pi_1^{({\rm static})} & = & \delta_{ij}\, \Pi_{ij}^{({\rm
static})} - 2 \,\frac{p_i p_j}{\vec{p}^2}\,\Pi_{ij}^{({\rm
static})},\nonumber\\
\Pi_2^{({\rm static})} & = & \Pi_{00}^{({\rm static})} -
\frac{p_i p_j}{\vec{p}^2}\, \Pi_{ij}^{({\rm static})},\nonumber\\ 
\Pi_3^{({\rm static})} & = & \frac{p_i p_j}{\vec{p}^2}\,
\Pi_{ij}^{({\rm static})}, \label{formfactorst}
\end{eqnarray}
while, in the long wave limit $(\vec{p}=0)$, we obtain
\begin{eqnarray}
\Pi_1^{({\rm long\ wave})} & = & - \Pi_{00}^{({\rm long\ wave})} +
\delta_{ij} \,\Pi_{ij}^{({\rm long\ wave})} - \frac{p_i
p_j}{\vec{p}^2}\, \Pi_{ij}^{({\rm long\ wave})}\,,\nonumber\\ 
\Pi_2^{({\rm long\ wave})} & = & - \Pi_{00}^{({\rm long\ wave})} +
\frac{p_i p_j}{\vec{p}^2} \,\Pi_{ij}^{({\rm long\ wave})}\,,\nonumber\\ 
\Pi_3^{({\rm long\ wave})} & = & \Pi_{00}^{({\rm long\
wave})}\;.\label{formfactorlw}
\end{eqnarray}
Note that $\frac{p_i p_j}{\vec{p}^2}\, \Pi_{ij} |_{\rm long\ 
wave}$ is well behaved. 

The calculation of the tadpole diagram is straightforward

\begin{equation}
 \Pi_{\mu \nu}^{({\rm tadpole})} = \frac{1}{2} (2 e^2
\delta_{\mu \nu}) \frac{1}{\beta} \sum_n \int \frac{d^2k}{(2 \pi)^2}
\frac{1}{k_0^2 + \omega_{\sigma}^2} 
=  e^2 \delta_{\mu \nu} \frac{\beta}{(2 \pi)^2} \sum_n \int
\frac{d^2k}{(2 \pi)^2}\,\frac{1}{n^2 + (\frac{\beta \omega_{\sigma}}{2
\pi})^2}, 
\end{equation}

\noindent where we have identified $k_0=\frac{2 \pi n }{\beta}$ and
$\omega_{\sigma} = (\vec{k}^2 + m_{\sigma}^2)^{1/2}$. The sum over the
Matsubara frequencies can be evaluated using 
\begin{equation}
\sum_n f(n) = - \pi \,\mbox{\rm Res} f(z) \cot \pi z,\label{sum}
\end{equation}
where the residues are calculated at the poles of the
function $f(z)$. Using this (as well as the periodic properties of the
trigonometric functions), we obtain 
\begin{equation}
\Pi_{\mu \nu}^{({\rm tadpole})} = \frac{e^2 \delta_{\mu
\nu}}{2} \int \frac{d^2 k}{(2 \pi)^2}\frac{1}{\omega_{\sigma}}
\coth(\frac{\beta \omega_{\sigma}}{2}) 
= \frac{e^2 \delta_{\mu \nu}}{2} \int \frac{d^2 k}{(2
\pi)^2}\frac{1}{\omega_{\sigma}} \left( 1 + \frac{2}{e^{\beta
\omega_{\sigma}}-1}\right). 
\end{equation}
The finite temperature contribution of the tadpole, therefore, follows to be
\begin{equation}
\Pi_{\mu \nu}^{({\rm tadpole})(T)} = e^2 \delta_{\mu \nu} \int
\frac{d^2 k}{(2 \pi)^2}\frac{1}{\omega_{\sigma}}\frac{1}{e^{\beta
\omega_{\sigma}-1}} =
-\delta_{\mu \nu} \,\frac{e^2}{2 \pi \beta} \ln(1 - e^{- \beta
m_{\sigma}}). \label{tadpole}
\end{equation}

As noted earlier, this diagram is independent of the external momentum
and, therefore, gives an analytic contribution. In fact, from the
definition in (\ref{formfactors}), we see that independent of the
static or  the long
wave limit, this diagram, eq. (\ref{tadpole}), leads to (we will
ignore the  superscript $T$
remembering all along that our interest is in the temperature
dependent part) 
\begin{equation}
\Pi_1^{({\rm tadpole})} =  0 = \Pi_2^{({\rm tadpole})},\quad
\Pi_3^{({\rm tadpole})} = - \frac{e^2}{2 \pi \beta}\,\ln(1 -
e^{- \beta m_{\sigma}}). 
\end{equation}
In fact, although the integrals, that we are interested in, can be
evaluated in closed form, for simplicity, let us consider the high
temperature limit, where we assume $T \gg m_i$ and yet is small
compared with the critical temperature where  symmetry may be restored 
(such a regime exists). In this limit, we have 
\begin{equation}
\Pi_1^{({\rm tadpole})} = 0 = \Pi_2^{({\rm tadpole})},\quad
\Pi_3^{({\rm tad pole})} \sim - \frac{e^2}{2 \pi \beta} \,\ln
\beta m_{\sigma}. \label{tadpole1}
\end{equation}

The rising sun diagram, on the other hand, does depend on the external
momentum and can, in principle, lead to a non-analytic contribution
for the parity conserving part of the self-energy 
\begin{equation}
 \Pi_{\mu \nu}^{({\rm rising\ sun})} = - (2 e m)^2
  \,\frac{1}{\beta} \sum_n \int \frac{d^2 k}{(2
  \pi)^2}\frac{\delta_{\mu \nu}\,(k_0^2 + \omega^2) + k_{\mu}k_{\nu}
  \,\frac{(k_0^2 + \omega^2 + \kappa^2)}{m^2}} {((k_0 + p_0)^2 +
  \omega_{\sigma}^2)(k_0^2+\omega_+^2)(k_0^2+\omega_-^2)}\label{risingsun} 
\end{equation}
where we have defined
\begin{equation}
\omega = (\vec{k}^2 + m^2)^{1/2},\quad
\omega_{\sigma} = ((\vec{k} + \vec{p})^2 + m_{\sigma}^2)^{1/2},\quad
\omega_{\pm} = (\vec{k}^2 + m_{\pm}^2)^{1/2}.
\end{equation}

Let us note here that the integrand in (\ref{risingsun}) involves
propagators  with
distinct masses. In such a case, it has been argued in \cite{arnold} that the
amplitude will be analytic at the origin in the energy-momentum
plane. More recently, it has been recognized that the non-analyticity
arises in the self-energy only if in some limits the integrand
develops double (or higher order) poles \cite{brandt}. When there are distinct
masses in the propagators, however, such a possibility cannot arise
and we will not expect a non-analyticity in the lowest order
terms. Nevertheless, let us evaluate the integral separately both in
the static limit as well as the long wave limit to understand this
further.

\subsection{Static Limit}

In this case, we set $p_0=0$ and evaluate the amplitude in
(\ref{risingsun}) as
$\vec{p}\rightarrow 0 $. For $\Pi_{00}^{({\rm rising\ sun})}$, we
obtain 
\begin{eqnarray}
\Pi_{00}^{({\rm rising\ sun})} &=& - 4\, e^2 m^2 \frac{1}{\beta}
\sum_n \int \frac{d^2 k}{(2 \pi)^2} \frac {(k_0^2 + \omega^2) + k_0^2
\frac{(k_0^2 + \omega^2 + \kappa^2)}{m^2}} {(k_0^2 +
\omega_{\sigma}^2)(k_0^2 +
\omega_+^2)(k_0^2 + \omega_-^2)}\nonumber \\ 
&=& - 4\, e^2 m^2 \frac{\beta^3}{(2 \pi)^4} \sum_n \int \frac{d^2
k}{(2 \pi)^2} \frac{(n^2 + (\frac{\beta \omega}{2 \pi})^2) + (\frac{2
\pi n}{\beta})^2\frac{(n^2 + (\frac{\beta}{2 \pi})^2(\omega^2 +
\kappa^2))}{m^2}}{ (n^2 + (\frac{\beta \omega_{\sigma}}{2 \pi})^2) (n^2 +
(\frac{\beta \omega_+}{2 \pi})^2) (n^2 + (\frac{\beta \omega_-}{2 \pi})^2) }. 
\end{eqnarray}

It is worth noting here that the  integrand has only simple poles
owing to the fact that the masses inside the loop are distinct. The
sum can  be evaluated, as before, using
eq. (\ref{sum}). Separating the temperature dependent part, we obtain the
value of this in the high temperature limit as $\vec{p} \rightarrow 0$
to be 
\begin{equation}
\lim_{\vec{p} \rightarrow 0} \Pi_{00}^{({\rm rising\ sun})(T)} =
\frac{e^2}{2 \pi \beta}\left[ 2 + \frac{4\, m^2\, (m^2 -
m_{\sigma}^2)}{(m_{\sigma}^2 - m_+^2)(m_{\sigma}^2 - m_-^2)}\, \ln
\frac{m_{\sigma}}{m_-} - \frac{4\, m^2\,(m^2 - m_+^2)}{(m_{\sigma}^2 -
m_+^2)(m_+^2- m_-^2)} \,\ln \frac{m_+}{m_-} \right]. \label{projst}
\end{equation} 
Here, we have used the high temperature limits of the integrals
\begin{eqnarray}
\int \frac{d^2 k}{(2 \pi)^2}\,\frac{1}{\omega} \frac{1}{e^{\beta \omega}-1}
& \longrightarrow & - \frac{1}{2 \pi \beta} \,\ln \beta m + {\cal
O}(\beta^0),\nonumber\\ 
\int \frac{d^2 k}{(2 \pi)^2}\, \frac{\omega}{e^{\beta \omega}-1} &
\longrightarrow &
\frac{2 \zeta(3)}{2 \pi \beta^3} - \frac{m^2}{4 \pi \beta} + {\cal
O}(\beta^0). 
\end{eqnarray}

The other projections can also be calculated, in the static limit, in
a similar manner. Without going into details, let us simply note the
high temperature limits of these quantities. 
\begin{eqnarray}
\lim_{\vec{p}\rightarrow 0} \delta_{ij} \Pi_{ij}^{({\rm rising\ 
sun})(T)} &\rightarrow & - \frac{e^2}{2 \pi \beta}\left[ - 4 \,\ln(\beta
m_{\sigma}) - \frac{4\,(2 m^4 - 3 m^2 m_+^2- m_+^2 \kappa^2 +
m_+^4)}{(m_{\sigma}^2 - m_+^2)(m_+^2 -
m_-^2)}\,\ln \frac{m_{\sigma}}{m_+}\right.\nonumber \\ 
& & \qquad + \left. \frac{4\,(2m^4 - 3 m^2m_-^2 - m_-^2 \kappa^2 +
m_-^4)}{(m_{\sigma}^2- m_-^2)(m_+^2
-m_-^2)}\,\ln \frac{m_{\sigma}}{m_-} + 2 \right],\nonumber\\ 
\lim_{\vec{p} \rightarrow 0} \frac{p_i p_j}{\vec{p}^2}\,
\Pi_{ij}^{({\rm rising\ sun})(T)} &= & \lim_{\vec{p}
\rightarrow 0} \frac{1}{2}\, \delta_{ij}\, \Pi_{ij}^{({\rm rising\ 
sun})(T)}. \label{proj1st}
\end{eqnarray}
The second of the above relations is, in fact, required by the Ward
identity of the theory and, consequently, our calculation is
consistent with the requirements of gauge invariance (BRS
invariance). It now follows, from eqs. (\ref{formfactorst}), (\ref{projst}) and
(\ref{proj1st}),  that in the static limit, as $\vec{p} \rightarrow 0$
(we drop the superscript $T$), 
\begin{equation}
\lim_{\vec{p} \rightarrow 0} \Pi_1^{({\rm rising\ sun})} = 0,
\end{equation}
and that the leading high temperature behavior of the other two form
factors is given by
\begin{eqnarray}
\lim_{\vec{p} \rightarrow 0} \Pi_2^{({\rm rising\ sun})} & = & - 2\,
\,\left(\frac{e^2}{2 \pi \beta} \,\ln  \beta m_{\sigma}\right) + {\cal
O}(\frac{1}{\beta}), \nonumber\\
\lim_{\vec{p} \rightarrow 0} \Pi_3^{({\rm rising\ sun})} & = &  2\,
\left(\frac{e^2}{2 \pi \beta} \,\ln  \beta m_{\sigma}\right) + {\cal
O}(\frac{1}{\beta}). 
\end{eqnarray}

Adding the contribution from the tadpole diagram,
eq. (\ref{tadpole1}), we see that, in the static limit, as $\vec{p}
\rightarrow 0$,
\begin{equation}
\Pi_1^{({\rm static})} = \Pi_1^{({\rm tadpole})} +
\Pi_1^{({\rm rising\ sun})} = 0,\label{Pi1}
\end{equation}
and that the leading high temperature behavior of the other two form
factors are given by
\begin{eqnarray} 
\Pi_2^{({\rm static})} & = & \Pi_2^{({\rm tadpole})} + \Pi_2^{({\rm
rising\ sun})} = - 2\,\left( \frac{e^2}{2 \pi \beta}\, \ln (\beta
m_{\sigma})\right),\nonumber\\ 
\Pi_3^{({\rm static})} & = & \Pi_3^{({\rm tadpole})} +
\Pi_3^{({\rm rising\ sun})} =  \left( \frac{e^2}{2 \pi \beta}
\,\ln (\beta m_{\sigma})\right).\label{static} 
\end{eqnarray}

\subsection{Long Wave Limit}

The form factors can also be calculated in a completely analogous
manner in the long wave limit, where we set $\vec{p}=0$ and look at the
amplitudes in the limit $p_0 \rightarrow 0$. Such a limit can be taken
only after the sum over the Matsubara frequencies have been carried
out and the external energies have been analytically continued to
Minkowski space. Let us indicate how this is done only in the case of
$\Pi_{00}$. 
\begin{equation}
\Pi_{00}^{({\rm rising\ sun})} = - 4 e^2 m^2 \,\frac{1}{\beta}
 \sum_n \int \frac{d^2 k}{(2 \pi)^2} \frac {(k_0^2 + \omega^2) +
 \frac{k_0^2\, (k_0^2 + \omega^2 +\kappa^2)}{m^2}}{ ((k_0 + p_0)^2 +
 \omega_{\sigma}^2) (k_0^2 + \omega_+^2) (k_0^2+\omega_-^2)}, 
\end{equation}
where $p_0=\frac{2 \pi\, l}{\beta}$ and (since $\vec{p}=0$) we have
now  defined 
$\omega_{\sigma}^2 = \vec{k}^2 + m_{\sigma}^2$. The sum can be evaluated
using eq. (\ref{sum}) as well as using the periodicity of trigonometric
functions. If we now analytically continue $p_0$ to Minkowski space
and look at the limit $p_0 \rightarrow 0$, then, the leading term is
identical to the leading term in the static limit. Therefore, the high
temperature limit leads to 
\begin{equation}
\lim_{p_0 \rightarrow 0} \Pi_{00}^{({\rm rising\ sun})(T)}
\rightarrow \frac{e^2}{2 \pi \beta}\left[2 + \frac{4 m^2 (m^2 -
m_{\sigma}^2)}{ (m_{\sigma}^2 - m_+^2) (m_{\sigma}^2 - m_-^2)}
\,\ln \frac{m_{\sigma}}{m_-} - \frac{4 m^2 (m^2 -
m_+^2)}{(m_{\sigma}^2 - m_+^2)(m_+^2 - m_-^2)}
\,\ln \frac{m_+}{m_-} \right]. 
\end{equation}

As we had alluded to earlier, the presence of distinct masses in the
propagator regulates the non-analyticity as a result of which the
lowest order term, in the long wave limit, is the same as in the
static limit \cite{arnold}.  This is also reflected in the other
calculations and yields 

\begin{eqnarray}
\lim_{p_0 \rightarrow 0} \delta_{ij} \,\Pi_{ij}^{({\rm rising\
sun})(T)} &\rightarrow & - \frac{e^2}{2 \pi \beta}\left[ - 4 \,\ln(\beta
m_{\sigma}) - \frac{4\,(2m^4 - 3 m^2 m_+^2 - m_+^2 \kappa^2 +
m_+^4)}{(m_{\sigma}^2- m_+^2)(m_+^2 -
m_-^2)}\,\ln \frac{m_{\sigma}}{m_+} \right.\nonumber \\ 
& & \qquad + \left.  \frac{4\,(2 m^4 - 3 m^2 m_-^2 - m_-^2 \kappa^2 +
m_-^4)}{(m_{\sigma}^2 - m_-^2)(m_+^2 -
m_-^2)}\,\ln \frac{m_{\sigma}}{m_-}  + 2 \right],\nonumber\\
\lim_{p_0 \rightarrow 0} \frac{p_i p_j}{\vec{p}^2}
\,\Pi_{ij}^{({\rm rising\ sun})(T)} & = & 
\lim_{p_0 \rightarrow 0} \frac{1}{2} \,\delta_{ij} \,\Pi_{ij}^{({\rm
rising\ sun})(T)}. 
\end{eqnarray}

However, even though the lowest order terms in the integrand are the
same in the two limits, the form factors are not. As can be seen from
eq. (\ref{formfactorlw}), in the long wave limit, we obtain (suppressing the
superscript $T$) the leading high temperature behaviors to be

\begin{eqnarray}
\lim_{p_0 \rightarrow 0} \Pi_1^{({\rm rising\ sun})} & = &  2\,
\left(\frac{e^2}{2 \pi \beta}\,\ln(\beta m_{\sigma}) \right) + {\cal
O}(\frac{1}{\beta}), \nonumber\\
\lim_{p_0 \rightarrow 0} \Pi_2^{({\rm rising\ sun})} & = &  2\,
\left(\frac{e^2}{2 \pi \beta}\,\ln(\beta m_{\sigma}) \right) + {\cal
O}(\frac{1}{\beta}), \nonumber\\
\lim_{p_0 \rightarrow 0} \Pi_3^{({\rm rising\ sun})} & = & {\cal
O}(\frac{1}{\beta}). 
\end{eqnarray}

Adding the contribution from the tadpole diagram,
eq. (\ref{tadpole1}),  the complete
form factors, in the long wave limit, have the leading high
temperature  behaviors, as $p_{0}\rightarrow 0$,
\begin{eqnarray}
\Pi_1^{({\rm long\ wave})} & = &  2\, \left(\frac{e^2}{2 \pi
\beta}\,\ln(\beta m_{\sigma}) \right),\nonumber\\ 
\Pi_2^{({\rm long\ wave})} & = &  2\, \left(\frac{e^2}{2 \pi
\beta}\,\ln(\beta m_{\sigma}) \right),\nonumber\\ 
\Pi_3^{({\rm long\ wave})} & = & -  \left(\frac{e^2}{2 \pi
\beta}\,\ln(\beta m_{\sigma}) \right).\label{longwave} 
\end{eqnarray}

\section{Discussion of results}

Our calculations are completely consistent with the known results
about loop diagrams with distinct masses in that the photon
self-energy is analytic in the lowest order
\cite{arnold,brandt}. However,  the form factors
are different in the static as well as the long wave limits. Beyond
the lowest order terms, however, we do not expect the distinct masses
in the propagators to lead to analytic results. This is already
evident in the parity violating part of the photon self-energy coming
from the rising sun diagram, where it is known that the leading high
temperature
behavior of the radiative correction to the Chern-Simons coefficient
is different in the static and the long wave limits \cite{alves}, 
\begin{eqnarray}
\lim_{\vec{p}\rightarrow 0}\,\Pi_{\rm odd}^{({\rm static})}  & = & 4
\kappa m^2  F(m_+, m_-,
m_{\sigma}) \left(\frac{e^2}{2 \pi \beta}\right),\nonumber\\ 
\lim_{p_{0}\rightarrow 0}\,\Pi_{\rm odd}^{({\rm long\ wave})} & = &
\frac{4 \kappa 
m^2}{(m_{\sigma}^2 - m_+^2)(m_{\sigma}^2-m_-^2)}\left(\frac{e^2}{2
\pi \beta}\,\ln \beta m_{\sigma} \right), 
\end{eqnarray}
where
\begin{equation}
F(m_+, m_-, m_{\sigma}) = \frac{m_+^2
\,\ln \frac{m_+}{m_{\sigma}}}{(m_{\sigma}^2-m_+^2)^2 (m_+^2 - m_-^2)
} - \frac{m_-^2 \,\ln \frac{m_-}{m_{\sigma}}}{(m_{\sigma}^2-m_-^2)^2
(m_+^2 - m_-^2) } + \frac{1}{2 (m_{\sigma}^2 -
m_+^2)(m_{\sigma}^2-m_-^2)}. 
\end{equation}

\begin{table}
\begin{center}
\begin{tabular}{lll}
Parameter & Static limit & Long wave Limit\\
  &  &  \\
$M^{2}_{1}$ & $\frac{e^2}{2\pi\beta}\,\ln\beta m_{\sigma}$ &
$\frac{e^2}{2\pi\beta}\,\ln\beta m_{\sigma}$ \\ 
$M^{2}_{2}$ & $- \frac{e^2}{2\pi\beta}\,\ln\beta m_{\sigma}$ &
$\frac{e^2}{2\pi\beta}\,\ln\beta m_{\sigma}$ \\ 
$\Pi_{\rm odd}$ &
$4\,\kappa\,m^2\,F(m_+,m_-,m_{\sigma})\,(\frac{e^2}{2\pi\beta})$ &
$\frac{4\,\kappa\,m^2\,(\frac{e^2}{2\pi\beta}\,\ln\beta m_{\sigma})}
{(m^{2}_{\sigma}-m^{2}_{+})(m^{2}_{\sigma}-m^{2}_{-})}$
\\ 
$M^{2}_{+}\simeq \Pi^{2}_{\rm odd}$ &
$16\,\kappa^2\,m^4\,F^2(m_+,m_-,m_{\sigma})\,(\frac{e^2}{2\pi\beta})^2$ &
$\frac{16\,\kappa^2\,m^4\,(\frac{e^2}{2\pi\beta}\,\ln\beta m_{\sigma})^2}
{(m^{2}_{\sigma}-m^{2}_{+})^2(m^{2}_{\sigma}-m^{2}_{-})^2}$
\\ 
$M^{2}_{-}$ & $(\ln\beta m_{\sigma})^2$ & ${\cal O}(1)$\\
\end{tabular}
\end{center}
\caption{Summary of results}
\end{table}

To understand the question of the screening length, let us tabulate
all the results that we know so far. Thus, we see, from table 1, that, at high
temperature,  the contribution of
$\Pi_{\rm odd}$ to $M_+^2$ is dominant and that $M_-^2$ is
negligible by comparison. Thus, for example, in the static limit, we
have (high $T$) 

\begin{eqnarray}
D_{00}^{({\rm static})} &=& Q_{00} \left[\frac{M_+^2 -
M_1^2}{M_+^2 - M_-^2}\frac{1}{\vec{p}^2 + M_+^2} - \frac{M_-^2 -
M_1^2}{M_+^2 - M_-^2}\frac{1}{\vec{p}^2 + M_-^2}\right]\nonumber \\ 
&\simeq& \frac{1}{\vec{p}^2 + M_+^2} +
\frac{M_1^2}{M_+^2}\;\frac{1}{\vec{p}^2}\nonumber \\ 
&=& \frac{1}{\vec{p}^2 + M_+^2} + \frac{ \,\pi\,\beta \,\ln(\beta
m_{\sigma})}{8 \,e^2\,\kappa^2 m^4\,
F^2(m_+,m_-,m_{\sigma})}\;\frac{1}{\vec{p}^2}\\ 
&\simeq& \frac{1}{\vec{p}^2 + M_+^2}.
\end{eqnarray} 

\noindent Namely, even though the propagator, $D_{00}$, has two poles, the
coefficient of the massless pole is negligible at high
temperature. Consequently, the propagator effectively has a single
pole and the screening length is related to  $M_+$ which is determined
by $\Pi_{\rm odd}$. It is also worth noting that the same massive pole
corresponds to the dominant term in the transverse part of $D_{ij}$ as well.

Although our discussion so far has been within the context of the
Abelian Chern-Simons Higgs theory, a similar behavior is also manifest
in the $2+1$ dimensional QED with a Chern-Simons term (where there is no
symmetry breaking). Let us note that in this theory, the tree level
propagator in a general covariant gauge has the form (in Euclidean
space) 

\begin{equation}
D_{\mu \nu}^{(0)} = \frac{1}{p^2 + \kappa^2} \left[(\delta_{\mu \nu} -
\frac{p_{\mu} \,p_{\nu}}{p^2})- \kappa \epsilon_{\mu \nu
\lambda}\,{p_{\lambda}\over p^{2}}\right] + \xi\,\frac{p_{\mu}
p_{\nu}}{(p^2)^2}, 
\end{equation} 
indicating a single massive pole. However, the complete two point
function, at finite temperature, can be parameterized as
(Conventionally, one identifies $\Pi_{1}=\Pi_{T}$ and
$\Pi_{2}=\Pi_{L}$. However, we will follow the notation of the earlier
section for consistency.)
\begin{equation}
\Gamma_{\mu \nu} = P_{\mu \nu} \,(p^2 + \Pi_1) + Q_{\mu \nu}\,(p^2 +
\Pi_2) + \epsilon_{\mu \nu \lambda} \,p_{\lambda} \,(\kappa +
\Pi_{\rm odd}) + \frac{1}{\xi}\,p_{\mu}p_{\nu},  
\end{equation}
leading to the complete propagator of the form
\begin{equation}
D_{\mu \nu} = \frac{1}{(p^2+M_+^2)(p^2+M_-^2)}\,\left[P_{\mu
\nu}\,(p^2+\Pi_2) + Q_{\mu \nu}\,(p^2+\Pi_1) - (\kappa +
\Pi_{\rm odd})\,\epsilon_{\mu \nu \lambda} \,p_{\lambda}\right] +
\frac{\xi\, p_{\mu} p_{\nu}}{(p^2)^2}\,. 
\end{equation}
Here, $M_{\pm}$ are the same as in (\ref{masses}) with $m=0=\Pi_{3}$.
We see that even though the tree level propagator has a single pole,
radiative corrections can generate two distinct poles in the
propagator, much like the Abelian Chern-Simons Higgs system. 

In the case of fermions, the masses inside the loop are identical (to
the fermion mass $M_{f}$). Therefore, we expect that the amplitudes
will be non-analytic \cite{brandt1}. For simplicity, we only list the leading
behavior of various quantities in the static limit, without giving any
technical details, which are quite standard. The radiative correction
to the Chern-Simons term has already been calculated for this theory
\cite{brandt1} 
and we have, in the static limit, (at high $T$)
\begin{equation}
\Pi_{\rm odd} = {e^{2}\over 8\pi}\,\beta M_{f}.
\end{equation}
Without giving details, we note that, in the static limit, the parity
conserving part of the self-energy yields
\begin{equation}
\lim_{\vec{p}\rightarrow 0}\,\Pi_{1} = \lim_{\vec{p}\rightarrow
0}\,\delta_{ij} \Pi_{ij}
 = 2e^{2} \int {d^{2}k\over
(2\pi)^{2}}\,{1\over \omega_{k}}\,{\partial\over
\partial\omega_{k}}\left({(\omega_{k}^{2}-m^{2})n_{F}(\omega_{k})\over
\omega_{k}}\right) = 0,\label{Pi1_1}
\end{equation}
where $n_{F}$ denotes the fermion distribution function and the
leading high  temperature behavior
\begin{equation}
\lim_{\vec{p}\rightarrow 0}\,\Pi_{2} = \lim_{\vec{p}\rightarrow
0}\,\Pi_{00}
 = {e^{2}\ln 2\over \pi\beta} + {\cal O} (\beta).
\end{equation}

Thus, we see that, in contrast to the Abelian Higgs model which we
have studied in detail in the earlier sections, here the contribution
of the parity violating part is negligible, at high temperature,
compared  with the parity
conserving part, namely, the roles of the parity conserving and the
parity violating parts appear to be reversed. In this case, it is easy
to calculate 
\begin{eqnarray}
M_{1}^{2} & = & \Pi_{1}\, =\, 0\, =\, M_{-}^{2},\nonumber\\
M_{+}^{2} & = & (\kappa + \Pi_{\rm odd})^{2} + \Pi_{2} \approx
{e^{2}\ln 2\over \pi\beta} + {\cal O}(\beta).
\end{eqnarray}
As a result, in this case, we have 
\begin{equation}
D_{00}^{({\rm static})} = Q_{00} \left[{M_{+}^{2}-M_{1}^{2}\over
M_{+}^{2}-M_{-}^{2}}\,{1\over \vec{p}^{2} + M_{+}^{2}} -
{M_{-}^{2}-M_{1}^{2}\over M_{+}^{2}-M_{-}^{2}}\,{1\over \vec{p}^{2} +
M_{-}^{2}}\right]
 = {1\over \vec{p}^{2} + M_{+}^{2}}.
\end{equation}
Once again, we see that, even though  {\em a priori} we would have
expected the existence of two poles, there is only one pole and that the
screening length is determined by $M_{+}$, which does not have any
leading  contribution from the parity violating form factor. This is
quite  different from the Abelian Higgs model
where it is the parity violating part of the form factor that
dominantly determines the screening length. We also note here that,
since $M_{1}=0 = M_{-}$ (and this is true to all orders as we will
argue shortly), the propagator in the static limit, has the 
behavior to all orders (in the Landau gauge)
\begin{equation}
D_{\mu\nu}^{({\rm static})} = P_{\mu\nu}\,{1\over \vec{p}^{2}} +
Q_{\mu\nu}\,{1\over \vec{p}^{2}+M_{+}^{2}} - {(\kappa + \Pi_{\rm
odd})\over \vec{p}^{2}
(\vec{p}^{2}+M_{+}^{2})}\,\epsilon_{\mu\nu\lambda} p_{\lambda}.
\end{equation}

The presence of the massless pole in the space-like components of the
propagator has already been observed in the non-Abelian theory
\cite{pisarski} and implies that static magnetic fields will not be screened in
this theory. This is reminiscent of the vanishing magnetic mass in QED
in $3+1$ dimensions and, in the present theory, this arises because
$\Pi_{1}=0$. (Namely, in this theory, the uniqueness of the screening
length  and the
absence of screening of static magnetic fields are
directly related.) Let us argue next that this holds true to all orders in
perturbation theory. Let us note that, gauge invariance (in QED)
implies that $p_{\mu}\Pi_{\mu\nu} = 0$, which in the static limit
gives
\begin{eqnarray}
p_{k}\Pi_{kj} & = & 0,\nonumber\\
{\rm or,}\quad \Pi_{ij} + p_{k} {\partial\Pi_{kj}\over \partial p_{i}} & =
& 0.
\end{eqnarray}
Assuming that the self-energy is analytic in the external momentum,
$\vec{p}$, (we note that, at finite temperature, amplitudes are
non-analytic in the external energy and momentum, but in the static
limit, they are analytic in the external momentum unless there are
infrared divergences), this implies upon using the symmetry properties
of the amplitude that the parity conserving part of
$\Pi_{ij}(0,\vec{p})\sim {\cal O}(\vec{p}^{2})$ 
so that $\Pi_{1}(0,0) = 0$. As we have mentioned, this formal argument
may be invalid when infrared divergences are present. While we have not
carried out any higher order calculation to verify this, we do not
expect infrared divergence to be a problem in the Abelian theory (the
infrared divergence is much more severe in the non-Abelian theory).
We would like to emphasize here that long range
correlations of static magnetic fields are well known in $3+1$
dimensional QED \cite{fradkin}. The interesting feature here is that
the  photon field
in $2+1$ dimensional theory is massive at the tree level because of the
Chern-Simons term and nonetheless a massless pole develops at the
loop level. Another interesting feature to note is that had we started
with a pure Chern-Simons theory \cite{hagen} (without the Maxwell
term)  interacting
with a fermion, the complete propagator would have the form (in the
static limit in the Landau gauge)
\begin{equation}
D_{\mu\nu}^{({\rm static})} = {1\over (\kappa +\Pi_{\rm
odd})^{2}\,\vec{p}^{2} + \Pi_{1}\Pi_{2}}\left[P_{\mu\nu}\,\Pi_{2} +
Q_{\mu\nu}\,\Pi_{1} - (\kappa + \Pi_{\rm
odd})\,\epsilon_{\mu\nu\lambda}p_{\lambda}\right].
\end{equation}
Since, $\Pi_{1}=0$, it follows that, in this case, there will be no
``$00$'' component of the gauge propagator. As a result, in this
theory, static
electric charges will not feel any force (which is, of course, true at
the tree level, but continues to hold at all loops), in addition to
static magnetic fields not being screened.

Finally, it is worth mentioning that even in the
Abelian-Chern-Simons-Higgs theory of section {\bf 2}, it is easy to
show using the Ward identities that $\Pi_{1} = 0$ to all orders in the
static limit (assuming no infrared divergence). This, however, does
not  result in a massless propagator
(unlike in the fermion theory) for the space-like indices. This
difference in the behavior of the two theories, namely, the fact that
in the fermion theory, static magnetic fields are not screened while
they are in the Abelian Higgs theory has a simple physical explanation. In the
fermion theory, there are no magnetic charges which can screen the
magnetic fields \cite{fradkin}, while the Abelian Higgs theory has
vortex  solutions which can achieve this.

In conclusion, we have systematically studied the question of
screening length in  $2+1$ dimensional Abelian theories with a
Chern-Simons term. We have shown that even though the starting gauge
propagator, in the Abelian Higgs theory has two poles, at finite
temperature, the coefficient of one of the poles becomes negligible
leading to a unique screening length that is related to the parity
violating part of the amplitude. In contrast, in a fermion theory, the
parity violating part is negligible and the screening length is
determined by the parity conserving part of the amplitude. In addition,
we have pointed out various other interesting features that arise in
these theories.

\vskip .5cm

One of us (A.D.) would like to thank Prof. J. Frenkel for many helpful
comments and discussions. We would like to dedicate this paper to the
memory of Van Cl\'{a}udio Calixto Alves. This work was supported in
part by US DOE Grant number DE-FG-02-91ER40685 and by CAPES, Brasil.



\begin{thebibliography}{99}

\bibitem{fradkin} E. Fradkin, Proc. Lebedev Phys. Inst. {\bf 29} (1965)
7.

\bibitem{kislinger} M. B. Kislinger and P. D. Morley, Phys. Rev. {\bf
D13} (1976) 2765.

\bibitem{weldon} H. A. Weldon, Phys. Rev. {\bf D26} (1982) 1394.

\bibitem{blaizot} J-P. Blaizot, E. Iancu and R. Parwani, Phys. Rev. {\bf
D52} (1995) 2543.

\bibitem{fetter-walecka} A. L. Fetter and J. D. Walecka, ``Quantum Theory
of Many Particle Systems'', McGraw-Hill, 1971.

\bibitem{weldon1} H. A. Weldon, Phys. Rev. {\bf D47} (1993) 594;
P. F. Bedaque and A. Das, Phys. Rev. {\bf D47} (1993) 601; A. Das and
M. Hott, Phys. Rev. {\bf D50} (1994) 6655.

\bibitem{deser} S. Deser, R. Jackiw and S. Templeton, Ann. Phys. {\bf
140} (1982) 372.

\bibitem{babu} K. S. Babu, A. Das and P. Panigrahi, Phys. Rev. {\bf D36}
(1987) 3725; E. Poppitz, Phys. Lett. {\bf B252} (1990) 417;
I. J. R. Aitchison, C. Fosco and J. Zuk, Ann. Phys. {\bf 242} (1995)
77.

\bibitem{khlebnikov} S. Yu. Khlebnikov, JETP Letters {\bf 51} (1990) 81;
V. Spiridonov, Phys. Lett. {\bf B247} (1990) 337; S. Yu. Khlebnikov
and M. Shaposnikov, Phys. Lett. {\bf B254} (1991) 148.

\bibitem{dunne} L. Chen, G. Dunne, K. Haller and E. Lim-Lombridas,
Phys. Lett. {\bf B348} (1995) 148.

\bibitem{khare} A. Khare, R. Mackenzie, P. Panigrahi and M. Paranjape,
Phys. Lett. {\bf B355} (1995) 236.

\bibitem{alves} V. S. Alves, A. Das, G. Dunne and S. Perez,
hep-th/0110160, to be published in Phys. Rev. D.

\bibitem{pisarski} R. Pisarski, Phys. Rev. {\bf D35} (1987) 664.

\bibitem{kapusta} J. Kapusta, ``Finite Temperature Field Theory'',
Cambridge University Press, 1989.

\bibitem{lebellac} M. Le Bellac, ``Thermal Field Theory'', Cambridge
University Press, 1996.

\bibitem{das} A. Das, ``Finite Temperature Field Theory'', World
Scientific, 1997.

\bibitem{arnold} P. Arnold, S. Vokos, P. Bedaque and A. Das,
Phys. Rev. {\bf D47} (1993) 4698.

\bibitem{brandt} F. T. Brandt, A. Das, J. Frenkel, J. C. Taylor,
hep-th/0112016, to be published in Phys. Rev. D.

\bibitem{brandt1} F. T. Brandt, A. Das and J. Frenkel, Phys. Rev. {\bf
D62} (2000) 085012.

\bibitem{hagen} C. R. Hagen, Ann. Phys. {\bf 157} (1984) 342.

\end{thebibliography}
\end{document}